\documentclass[11pt]{article}

\usepackage[margin=1in]{geometry}
\usepackage{amsmath}
\usepackage{booktabs}
\usepackage{hyperref}
\usepackage{enumitem}
\usepackage{xcolor}
\usepackage{array}

\hypersetup{
  colorlinks=true,
  linkcolor=blue,
  citecolor=blue,
  urlcolor=blue
}

\setlist[itemize]{leftmargin=1.5em}
\newcommand{\qwenawq}{Qwen2.5\nobreakdash-72B\nobreakdash-Instruct\nobreakdash-AWQ}
\newcommand{\qwenbase}{Qwen2.5\nobreakdash-72B}

\title{Sustained 70B-Class AWQ Inference on a Single NVIDIA L20:\\
Throughput, Stability, Energy, and Quality Characterization}

\author{
Yin Li\\
\texttt{https://github.com/Kevin-Li-2025/llm-quant-bench}
}

\date{Technical Report Draft: May 20, 2026}

\begin{document}
\maketitle

\begin{abstract}
Serving 70B-class open-weight language models is usually associated with
80GB accelerators, tensor-parallel multi-GPU systems, or vendor-managed
inference profiles. This report asks a narrower systems question: can a single
NVIDIA L20 48GB GPU sustain a useful 70B-class quantized serving workload, and
what are the throughput, latency, stability, energy, and quality boundaries of
that configuration? We evaluate \qwenawq{} served with vLLM
0.8.5.post1 and AWQ Marlin on one L20. Under a fixed workload of approximately
512 input tokens and 256 output tokens, a 24-hour concurrency-10 soak completed
36,740/36,740 requests with no request failures and no vLLM CUDA OOM, traceback,
or killed-process signatures. The system sustained 108.84 output tokens/s, with
p95 time-to-first-token (TTFT) of 6.61s and p95 end-to-end latency of 23.54s.
GPU-board power sampled through \texttt{nvidia-smi} produced an estimated
7.92 kWh over the run, corresponding to 0.330 output tokens/J and 1.008 total
tokens/J. Repeated fixed-shape runs at concurrency 1, 4, 8, and 16 completed
12/12 runs successfully; the concurrency-16 condition averaged
127.22 $\pm$ 12.68 output tokens/s over three runs, matching the earlier
single-run screening result of 127.70 output tokens/s. The same AWQ endpoint
also produced absolute quality scores of 0.8130 on MMLU, 0.8309 on CMMLU, and
0.8082 on GSM8K, plus 80/80 MT-Bench answer generations and a 60-item 8K
LongBench subset.

The evidence supports a specific claim: a carefully configured single L20 can
serve \qwenawq{} as a throughput-oriented 70B-class endpoint under
the tested fixed-shape workload. It does not prove lossless AWQ quality
retention, low-latency interactive serving, broad production SLA coverage, or
equivalence to a BF16/FP16 baseline. BF16/FP16 baseline retention and MT-Bench
judge scoring remain blocked until external or multi-GPU baseline and judge
endpoints are available.
\end{abstract}

\section{Executive Summary}

The central result is operational: a single NVIDIA L20 can repeatedly and
stably serve a 70B-class AWQ model when the serving profile is designed around
short-context, throughput-oriented batching. Table~\ref{tab:headline} gives the
headline measurements.

\begin{table}[h]
\centering
\small
\begin{tabular}{>{\raggedright\arraybackslash}p{0.38\linewidth}
                >{\raggedright\arraybackslash}p{0.50\linewidth}}
\toprule
Measurement & Result \\
\midrule
Model & \qwenawq{} \\
Runtime & vLLM 0.8.5.post1, AWQ Marlin \\
GPU & 1x NVIDIA L20, 46GB visible VRAM \\
Fixed-shape workload & approx. 512 input / 256 output tokens \\
24h soak success & 36,740/36,740 requests \\
24h soak throughput & 108.84 output tokens/s \\
24h soak p95 TTFT / latency & 6.61s / 23.54s \\
24h GPU-board energy & 7.92 kWh \\
24h output tokens/J & 0.330 \\
Repeated c16 throughput & 127.22 $\pm$ 12.68 output tokens/s \\
Repeated c16 p95 TTFT / latency & 11.91s $\pm$ 0.15s / 34.95s $\pm$ 1.00s \\
Repeated run success & 12/12 runs, 100\% success \\
Quality, absolute AWQ & MMLU 0.8130, CMMLU 0.8309, GSM8K 0.8082 \\
Long-context subset & 60/60 LongBench 8K subset requests completed \\
Explicit non-claim & No BF16/FP16 retention or MT-Bench judge score yet \\
\bottomrule
\end{tabular}
\caption{Headline results. Confidence intervals are two-sided 95\% intervals over
three repeated run-level summaries using Student $t$ critical values.}
\label{tab:headline}
\end{table}

This report should be read as an empirical systems characterization, not as a
new quantization algorithm paper. The contribution is the measurement package:
serving configuration, fixed-shape throughput, repeated-run variance, 24-hour
stability, GPU-board energy, absolute quality checks, and explicit blockers for
stronger quality-retention claims.

\section{Scope and Non-Claims}

The supported claim is:

\begin{quote}
A single NVIDIA L20 can sustain \qwenawq{} serving under a fixed
approximately 512 input / 256 output token workload, with 24-hour stability at
concurrency 10 and repeated short-run throughput near 127 output tokens/s at
concurrency 16.
\end{quote}

The following claims are not supported by the current evidence:

\begin{itemize}
  \item AWQ is lossless relative to BF16/FP16.
  \item The configuration is a low-latency chat endpoint.
  \item The result generalizes to all 70B models, all prompts, or all
        quantization formats.
  \item The 8K LongBench subset is an official LongBench leaderboard result.
  \item MT-Bench has a judge score.
  \item Single-L20 AWQ generally replaces multi-GPU BF16/FP16 inference.
\end{itemize}

The non-claims are important. A technical report is only useful if the
boundaries are as concrete as the positive result.

\section{Introduction}

The practical value of 70B-class open-weight language models is clear: they are
strong enough for many enterprise and research workloads while remaining
self-hostable. The deployment difficulty is that most 70B serving discussions
assume 80GB GPUs, multi-GPU tensor parallelism, or vendor-optimized inference
profiles. Many organizations, however, have access to smaller datacenter
accelerators such as the NVIDIA L20. A 48GB-class GPU can hold 4-bit 70B weights,
but static model fit does not imply sustained serving capacity. KV-cache
allocation, prompt shape, output length, batching policy, quantization kernels,
thermal behavior, and tail latency determine whether the system is useful.

This report studies one concrete question:

\begin{quote}
What can one NVIDIA L20 actually sustain when serving a 70B-class AWQ model
under fixed, reproducible workloads?
\end{quote}

The answer is empirical. We do not propose a new model, quantization method, or
serving algorithm. Instead, we characterize a constrained deployment point that
is often discussed informally but rarely reported with day-long stability,
confidence intervals, energy estimates, and quality checks.

\paragraph{Contributions.}

\begin{itemize}
  \item We report a 24-hour concurrency-10 fixed-shape soak of
        \qwenawq{} on one NVIDIA L20, including success rate,
        TTFT, latency, output throughput, request throughput, GPU-board energy,
        and log-based OOM/error checks.
  \item We repeat the fixed-shape workload at concurrency 1, 4, 8, and 16 with
        three runs per condition, reporting run-level means and 95\% confidence
        intervals.
  \item We report absolute AWQ quality measurements on MMLU, CMMLU, GSM8K,
        MT-Bench answer generation, and an 8K LongBench subset
        \cite{mt_bench_arena,longbench}.
  \item We compare the measured operating point against public Qwen 2.5 72B Q4
        serving numbers, Qwen official speed benchmarks, and NVIDIA NIM
        supported-profile guidance.
  \item We document the exact blockers for BF16/FP16 baseline retention,
        MT-Bench judge scoring, runtime ablation, and quantization ablation.
\end{itemize}

\section{Background}

\subsection{Why 70B on 48GB Is Nontrivial}

Weight-only 4-bit quantization can reduce the weight footprint of a 70B-class
model enough to fit in a 48GB GPU. Serving still requires more than weights:
KV cache, runtime buffers, scheduling state, tokenization overhead, and memory
fragmentation all consume capacity. Increasing context length or concurrency can
push the same model from stable to OOM. Therefore, ``runs on one L20'' is not a
single claim. It can mean a one-request smoke test, an interactive endpoint, a
high-throughput fixed-shape service, or a long-running production workload. This
report focuses on the latter two under explicitly bounded shapes.

\subsection{vLLM, Continuous Batching, and AWQ Marlin}

vLLM is a high-throughput LLM serving engine built around efficient KV-cache
management and continuous batching. The PagedAttention work reports large
throughput gains from better memory management during LLM serving
\cite{vllm_pagedattention}. AWQ is a post-training weight-only quantization
method that protects salient channels identified through activation statistics
\cite{awq}. Marlin is an optimized mixed-precision FP16-by-INT4 inference kernel
for autoregressive LLM inference \cite{marlin}. The measured system uses vLLM's
AWQ Marlin path, \texttt{awq\_marlin}. This matters because a 4-bit checkpoint
without an optimized serving kernel is a different deployment configuration.

\subsection{Serving Metrics}

We follow the serving metric vocabulary used by vLLM and NVIDIA GenAI-Perf:
time-to-first-token (TTFT), end-to-end request latency, inter-token latency
(ITL), time per output token (TPOT), request throughput, and output token
throughput \cite{vllm_bench_serve,genai_perf_docs}. In a continuously batched
system, aggregate server output tokens/s can be much higher than per-request
decode speed. This is central to the result: the 24-hour concurrency-10 run
produces 108.84 aggregate output tokens/s while individual request decode speed
is roughly 11--15 tokens/s.

\section{Experimental Setup}

\subsection{Hardware and Software}

\begin{table}[h]
\centering
\small
\begin{tabular}{ll}
\toprule
Component & Configuration \\
\midrule
GPU & 1x NVIDIA L20 \\
Visible GPU memory & 46,068 MiB reported by \texttt{nvidia-smi} \\
Model & Qwen/\qwenawq{} \\
Runtime & vLLM 0.8.5.post1 \\
Quantization path & AWQ Marlin \\
Serving API & OpenAI-compatible \texttt{/v1/chat/completions} \\
Primary context profile & \texttt{max\_model\_len=1024} \\
Power sampling & \texttt{nvidia-smi}, approximately every 10 seconds \\
\bottomrule
\end{tabular}
\caption{Primary test configuration.}
\label{tab:setup}
\end{table}

The primary service command was:

\begin{verbatim}
vllm serve /home/USER/models/Qwen2.5-72B-Instruct-AWQ \
  --host 0.0.0.0 --port 8001 \
  --served-model-name qwen72b-awq-l20 \
  --quantization awq_marlin \
  --dtype half \
  --max-model-len 1024 \
  --gpu-memory-utilization 0.98 \
  --max-num-seqs 48 \
  --max-num-batched-tokens 4096 \
  --enforce-eager \
  --swap-space 1 \
  --disable-log-requests \
  --trust-remote-code
\end{verbatim}

This is a short-context throughput profile. Separate 4096- and 8192-context
single-concurrency checks were run, but the main throughput claim is for the
1024-context service profile.

\subsection{Workload}

The fixed-shape benchmark was designed to be close to public serving tables:

\begin{itemize}
  \item 128 unique prompts.
  \item 498 raw tokenizer prompt tokens.
  \item Approximately 527 server-side prompt tokens after chat formatting.
  \item Fixed output length of 256 tokens.
  \item \texttt{max\_tokens=256}, \texttt{min\_tokens=256},
        \texttt{ignore\_eos=true}, \texttt{temperature=0}.
  \item Streaming responses with \texttt{stream\_options.include\_usage=true}.
\end{itemize}

The 24-hour soak reused the same workload at concurrency 10. Repeated short runs
used concurrency 1, 4, 8, and 16 with three repeats per condition.

This benchmark is MLPerf-inspired in the sense that it fixes the system under
test, workload shape, quality checks, and tail-latency measurements, but it is
not an MLPerf submission. It does not use MLPerf LoadGen, a Poisson server
scenario, an audited latency constraint, or an official quality target. The
numbers should therefore be interpreted as repo-native fixed-shape serving
measurements.

\subsection{Energy and Confidence Intervals}

GPU-board energy is computed from sampled board power:

\begin{align}
E_\mathrm{J} &= \int P(t)\,dt, \\
\mathrm{output\_tok/J} &= \frac{\sum_i \mathrm{output\_tokens}_i}{E_\mathrm{J}}, \\
\mathrm{total\_tok/J} &=
  \frac{\sum_i(\mathrm{prompt\_tokens}_i+\mathrm{output\_tokens}_i)}{E_\mathrm{J}}.
\end{align}

The implementation uses trapezoidal integration over the power trace. These are
GPU-board energy estimates, not wall-power measurements.

Repeated-run confidence intervals use run-level summaries:

\begin{align}
\bar{x} &= \frac{1}{n}\sum_{i=1}^{n}x_i, \\
\mathrm{CI}_{95} &= \bar{x} \pm t_{0.975,n-1}\frac{s}{\sqrt{n}}.
\end{align}

For the reported repeated runs, $n=3$ per concurrency.

\section{Serving Results}

\subsection{Fixed-Shape Screening Sweep}

Table~\ref{tab:fixed_shape_sweep} shows the initial fixed-shape concurrency
sweep. Throughput improves through concurrency 16 and then regresses at
concurrency 24, where tail latency also increases sharply.

\begin{table}[h]
\centering
\small
\begin{tabular}{rrrrrrrr}
\toprule
Concurrency & Requests & Success & p95 TTFT & p95 Lat. & Out tok/s & Req/s & p05 decode \\
\midrule
1  & 8   & 100\% & 0.76s  & 15.88s & 16.21  & 0.063 & 16.92 \\
4  & 32  & 100\% & 2.94s  & 18.63s & 57.10  & 0.223 & 14.99 \\
8  & 64  & 100\% & 6.06s  & 22.20s & 93.38  & 0.365 & 12.13 \\
10 & 80  & 100\% & 6.71s  & 23.69s & 108.70 & 0.425 & 11.26 \\
16 & 128 & 100\% & 11.01s & 36.29s & 127.70 & 0.499 & 7.66 \\
24 & 120 & 100\% & 37.26s & 64.08s & 120.51 & 0.471 & 5.25 \\
\bottomrule
\end{tabular}
\caption{Initial fixed-shape sweep, approximately 512 input tokens and 256 output
tokens. Concurrency 16 is the highest-throughput screened point; concurrency 24
is beyond the useful peak for this profile.}
\label{tab:fixed_shape_sweep}
\end{table}

\subsection{Repeated Fixed-Shape Runs}

The c1/c4/c8/c16 conditions were repeated three times each. Table~\ref{tab:ci}
reports run-level means and two-sided 95\% confidence intervals.

\begin{table}[h]
\centering
\small
\begin{tabular}{rrrrrrr}
\toprule
Conc. & Runs & Success & Out tok/s & Req/s & p95 TTFT & p95 Lat. \\
\midrule
1  & 3 & 100\% $\pm$ 0.00\% & 16.57 $\pm$ 1.21  & 0.065 $\pm$ 0.005 & 0.54s $\pm$ 1.90s  & 15.66s $\pm$ 1.94s \\
4  & 3 & 100\% $\pm$ 0.00\% & 55.75 $\pm$ 3.54  & 0.218 $\pm$ 0.014 & 3.01s $\pm$ 0.17s  & 18.70s $\pm$ 0.16s \\
8  & 3 & 100\% $\pm$ 0.00\% & 93.26 $\pm$ 0.23  & 0.364 $\pm$ 0.001 & 6.05s $\pm$ 0.06s  & 22.16s $\pm$ 0.05s \\
16 & 3 & 100\% $\pm$ 0.00\% & 127.22 $\pm$ 12.68 & 0.497 $\pm$ 0.050 & 11.91s $\pm$ 0.15s & 34.95s $\pm$ 1.00s \\
\bottomrule
\end{tabular}
\caption{Repeated fixed-shape runs. All twelve runs completed successfully with
no vLLM OOM/error signatures.}
\label{tab:ci}
\end{table}

The repeated concurrency-16 mean, 127.22 output tokens/s, matches the earlier
single-run result of 127.70 output tokens/s. The concurrency-8 result is
especially tight at 93.26 $\pm$ 0.23 output tokens/s. The concurrency-16
throughput interval is wider because two runs produced roughly 130 output
tokens/s and one run produced 121.32 output tokens/s. Tail latency remains the
main cost of the high-throughput profile.

\subsection{Twenty-Four-Hour Soak}

Table~\ref{tab:soak} reports the 24-hour concurrency-10 soak. The run completed
normally, and log inspection found zero CUDA OOM, OutOfMemory, Traceback, ERROR,
or Killed signatures.

\begin{table}[h]
\centering
\small
\begin{tabular}{lr}
\toprule
Metric & Value \\
\midrule
Duration & 86,412.60s \\
Total requests & 36,740 \\
Successful requests & 36,740 \\
Failed requests & 0 \\
Success rate & 100\% \\
Prompt tokens & 19,361,980 \\
Output tokens & 9,405,440 \\
Total tokens & 28,767,420 \\
Output throughput & 108.84 output tok/s \\
Request throughput & 0.425 req/s \\
p95 TTFT & 6.61s \\
p95 latency & 23.54s \\
p05 per-request decode speed & 11.26 tok/s \\
OOM/error signatures & 0 \\
\bottomrule
\end{tabular}
\caption{Twenty-four-hour fixed-shape soak at concurrency 10. The 24-hour
throughput, 108.84 output tokens/s, matches the short concurrency-10 screening
result of 108.70 output tokens/s.}
\label{tab:soak}
\end{table}

The agreement between the short screening run and the day-long run is one of the
strongest results in the report. A short result can be a transient peak; a
24-hour run at the same throughput with zero failures is evidence of sustained
stability under this exact workload.

\subsection{GPU-Board Energy}

\begin{table}[h]
\centering
\small
\begin{tabular}{lr}
\toprule
Metric & Value \\
\midrule
Power samples & 8,618 \\
Power trace duration & 86,410.0s \\
Average GPU board power & 330.15W \\
GPU board energy & 28,528,468J \\
GPU board energy & 7.92 kWh \\
Output tokens/J & 0.330 \\
Total tokens/J & 1.008 \\
Joules/output token & 3.033 \\
Joules/total token & 0.992 \\
\bottomrule
\end{tabular}
\caption{GPU-board energy estimate from \texttt{nvidia-smi} power samples. This
is not a full-system wall-power measurement.}
\label{tab:energy}
\end{table}

The energy result is useful for regression testing and rough deployment
planning. It should not be used as a full data-center energy estimate because
CPU, memory, fans, power-supply losses, and cooling are outside the measurement
scope.

\section{Quality Characterization}

This report does not yet include BF16/FP16-vs-AWQ retention. It does include
absolute quality measurements from the AWQ candidate. These are useful for
detecting severe quality collapse and for documenting the measured candidate,
but they do not prove that AWQ is lossless.

\begin{table}[h]
\centering
\small
\begin{tabular}{lrrrrr}
\toprule
Evaluation & Context & Items & OK & Failed & Score \\
\midrule
MMLU & 1024 & 14,042 & 14,039 & 3 & 0.8130 \\
CMMLU & 1024 & 11,582 & 11,582 & 0 & 0.8309 \\
GSM8K & 1024 & 1,319 & 1,319 & 0 & 0.8082 \\
MT-Bench generation & 1024 & 80 & 80 & 0 & pending judge \\
LongBench subset, token-F1 & 8192 & 60 & 60 & 0 & 0.2038 \\
LongBench v1 task metrics & 8192 & 60 & 60 & 0 & 16.16\% \\
\bottomrule
\end{tabular}
\caption{Absolute AWQ candidate quality measurements. MT-Bench currently reports
answer generation only, not a judge score. LongBench is a 60-item 8K subset, not
an official leaderboard run.}
\label{tab:quality}
\end{table}

The three MMLU failures were context-limit validation failures from prompts that
exceeded the 1024-token short-context service profile, not CUDA OOMs. The
LongBench v1 task-metric postprocess scored the 60 generated samples with
task-specific metrics: 38.67\% on \texttt{multifieldqa\_en}, 30.53\% on
\texttt{hotpotqa}, 12.90\% on \texttt{multi\_news}, 12.08\% on
\texttt{gov\_report}, 2.76\% on \texttt{lcc}, and 0.00\% on
\texttt{passage\_count}. These numbers are stricter than the earlier
single-token-F1 summary, but they remain a subset result.

\subsection{Blocked Retention and Judge Results}

We ran a preflight for the two missing quality claims. Table~\ref{tab:blockers}
summarizes the result.

\begin{table}[h]
\centering
\small
\begin{tabular}{lll}
\toprule
Item & Status & Missing prerequisites \\
\midrule
BF16/FP16 baseline retention & Blocked &
\texttt{BASELINE\_BASE\_URL}, \texttt{BASELINE\_MODEL} \\
MT-Bench judge score & Blocked &
\texttt{JUDGE\_BASE\_URL}, \texttt{JUDGE\_MODEL}, \texttt{JUDGE\_API\_KEY} \\
\bottomrule
\end{tabular}
\caption{Quality-retention and MT-Bench judge blockers as of the latest
preflight.}
\label{tab:blockers}
\end{table}

The remote host currently stages only AWQ checkpoint directories under
\texttt{/home/USER/models}. A local 72B BF16/FP16 baseline is not feasible on
one L20: 72B BF16/FP16 weights require roughly 144GB before KV cache and runtime
overhead, while the host exposes about 46GB of GPU memory and about 63GB of
free disk. We also intentionally avoid judging MT-Bench with the AWQ candidate
itself, because self-judging would not be a valid MT-Bench judge score.

\section{Long-Context Capacity Checks}
\label{sec:long_context}

\begin{table}[h]
\centering
\small
\begin{tabular}{rrrrrrr}
\toprule
Context & Prompt tokens & Concurrency & Success & p95 TTFT & Out tok/s & OOM \\
\midrule
4096 & 3,875 & 1 & 5/5 & 5.28s & 10.02 & No \\
8192 & 7,514 & 1 & 3/3 & 11.03s & 6.16 & No \\
\bottomrule
\end{tabular}
\caption{Single-concurrency long-context capacity checks. These are not
high-throughput long-context serving results.}
\label{tab:long_context}
\end{table}

The 4K and 8K checks show that the model can handle longer prompts at
concurrency 1 without OOM. They do not justify a high-concurrency long-context
claim. Long-context throughput requires a separate service profile, workload
matrix, and soak.

\section{External Positioning}

Table~\ref{tab:external} compares the result with public numbers. These
comparisons are directional because serving throughput depends on model variant,
quantization format, kernel, scheduler, prompt length, output length,
concurrency, sampling settings, and measurement method.

\begin{table}[h]
\centering
\scriptsize
\setlength{\tabcolsep}{3pt}
\begin{tabular}{>{\raggedright\arraybackslash}p{0.14\linewidth}
                >{\raggedright\arraybackslash}p{0.13\linewidth}
                >{\raggedright\arraybackslash}p{0.25\linewidth}
                >{\raggedright\arraybackslash}p{0.25\linewidth}
                r}
\toprule
Source & Hardware & Model / Quant & Shape & Result \\
\midrule
This work & 1x L20 48GB & \qwenbase{} AWQ Marlin & 512/256, c10, 24h & 108.84 \\
This work & 1x L20 48GB & \qwenbase{} AWQ Marlin & 512/256, c16, repeated & 127.22 $\pm$ 12.68 \\
GigaGPU Apr. 2026 & 1x RTX 3090 & Qwen 2.5 72B Q4 & 512/256, c10 & 32 \\
GigaGPU Apr. 2026 & 1x RTX 5090 & Qwen 2.5 72B Q4 & 512/256, c10 & 58--82 \\
GigaGPU Apr. 2026 & 1x RTX 6000 Pro & Qwen 2.5 72B Q4 & 512/256, c10 & 45 \\
Qwen official & 1x A100 80GB & \qwenbase{} AWQ, Transformers & input 1, output 2048, batch 1 & 11.50 \\
Qwen official & 2x A100 80GB & \qwenbase{} AWQ, vLLM & input 1, output 2048, batch 1 & 44.30 \\
Qwen official & 2x A100 80GB & \qwenbase{} AWQ, vLLM & input 30720, output 2048, batch 1 & 30.02 \\
\bottomrule
\end{tabular}
\caption{Directional comparison. The GigaGPU rows are the closest public
512-input/256-output/10-concurrent shape. Qwen official rows are batch-1 speed
references and should not be directly compared to aggregate concurrency-10 or
concurrency-16 throughput.}
\label{tab:external}
\end{table}

The closest public table is GigaGPU's April 2026 benchmark, which reports
512-token prompts, 256-token generations, and 10 concurrent users for Qwen 2.5
72B Q4 \cite{gigagpu_2026}. Under that shape, this single-L20 AWQ Marlin result
is above the published single-GPU figures in that table. This does not prove
that L20 is generally faster than those GPUs; it shows that this specific
vLLM/AWQ-Marlin configuration is competitive under the tested workload.

The model itself is not new. \qwenbase{}-Instruct is documented by the Qwen
team's model card and technical report \cite{qwen25_model_card,qwen25_report}.
Qwen's official speed benchmark is useful context for batch-1 and long-output
operating points, but it measures different shapes from the aggregate serving
workloads studied here \cite{qwen25_speed_benchmark}.

NVIDIA NIM's supported-model table is also relevant. Its optimized Qwen2.5 72B
Instruct FP8 L20 profiles list 4 or 8 L20 GPUs, while A100 SXM BF16 profiles
also use multi-GPU configurations depending on profile \cite{nvidia_nim_supported}.
Our result demonstrates a different deployment point: single-L20 AWQ Marlin
outside that conservative FP8 profile matrix.

\section{Discussion}

\subsection{The Result Is Throughput-Oriented}

The service is not a low-latency chat profile. At concurrency 16, the repeated
run mean p95 TTFT is 11.91s and p95 end-to-end latency is 34.95s for fixed
256-token outputs. This is useful for batch generation, asynchronous agents,
offline summarization, internal processing, and throughput-oriented serving. It
is not ideal for interactive applications that require sub-second first tokens.

\subsection{Why Repetition Matters}

The repeated runs strengthen the report because they show that the screening
numbers were not one-off peaks. Concurrency 8 is highly stable across three
runs. Concurrency 16 has wider throughput variance, but the mean remains close
to the initial single-run result and all runs complete successfully. The 24-hour
soak provides a different kind of evidence: long-run operational stability at
the more balanced concurrency-10 point.

\subsection{What Would Make This a Stronger Research Paper}

The report becomes closer to a full research paper if it answers comparative
questions rather than only characterizing one deployment point. The highest
priority additions are:

\begin{itemize}
  \item BF16/FP16 baseline quality retention using the same prompts, datasets,
        decoding settings, and scorers.
  \item MT-Bench judge scoring with an external judge model and answer-order
        swapping.
  \item Runtime ablation: vLLM versus SGLang versus llama.cpp/GGUF.
  \item Quantization ablation: AWQ versus GPTQ versus FP8 where compatible.
  \item Multiple models: Llama 70B, DeepSeek distill 70B, and Qwen3-class 70B
        checkpoints where available.
\end{itemize}

\section{Reporting Checklist Against Strong Technical Reports}

Strong modern technical reports tend to combine a narrow central claim, complete
experimental context, broad evaluations, clear failure analysis, and explicit
release artifacts. GPT-4 emphasizes capability, limitation, and safety
evaluation while withholding some architecture details \cite{gpt4_report}.
Llama 3 and Gemini 1.5 provide broad evaluation suites across model variants and
modalities \cite{llama3_report,gemini15_report}. DeepSeek-V3 highlights
architecture, training stability, cost, and inference efficiency
\cite{deepseekv3_report}. MLPerf Inference is not an LLM model report, but it
sets a useful benchmark-reporting standard: fixed scenarios, quality targets,
latency bounds, submission review, and reproducible result artifacts
\cite{mlperf_inference}.

Table~\ref{tab:reporting_checklist} audits this report against that standard.
The purpose is to prevent overclaiming and to make the remaining work obvious.

\begin{table}[h]
\centering
\scriptsize
\setlength{\tabcolsep}{3pt}
\begin{tabular}{>{\raggedright\arraybackslash}p{0.22\linewidth}
                >{\raggedright\arraybackslash}p{0.22\linewidth}
                >{\raggedright\arraybackslash}p{0.22\linewidth}
                >{\raggedright\arraybackslash}p{0.22\linewidth}}
\toprule
Dimension & Strong-report expectation & Current status & Upgrade needed \\
\midrule
Claim framing & One central claim with explicit scope and non-claims &
Strong: supported claim and non-claims are stated early &
Keep the claim narrow in title, abstract, and README \\
Method details & Hardware, software, workload, decoding, and metrics &
Strong for the measured vLLM/AWQ setup &
Add driver/CUDA exact versions and wall-power instrumentation \\
Statistical rigor & Repeated runs, uncertainty, and tail metrics &
Moderate: 3 repeats per short condition plus 24h soak &
Run more repeats across days or machines \\
Quality evidence & Matched baselines and benchmark-specific scoring &
Partial: MMLU/CMMLU/GSM8K absolute scores, MT-Bench generation, LongBench subset &
Add BF16/FP16 retention, MT-Bench judge, and official LongBench runs \\
Comparative evidence & Controlled ablations against alternatives &
Weak: public directional comparisons only &
Add vLLM/SGLang/llama.cpp and AWQ/GPTQ/FP8 ablations \\
Failure analysis & OOM, errors, tails, and boundary conditions &
Moderate: OOM/error scan and c24 tail-latency regression are reported &
Add explicit failure-point sweeps for context and concurrency \\
Energy/cost & Measurement scope and cost model &
Partial: GPU-board energy only &
Measure wall power and convert to cost/token \\
Artifacts & Code, configs, raw outputs, and source package &
Strong: repo, commands, generated PDF, and arXiv source package &
Publish raw logs/checksums or a reproducible artifact archive \\
Narrative/figures & A few plots that make the result visually obvious &
Weak: mostly tables &
Add throughput-latency curves, power trace, quality summary figures, and an
MLPerf-style server scenario if comparability becomes the goal \\
\bottomrule
\end{tabular}
\caption{Self-audit against common expectations in high-quality technical
reports and benchmark reports.}
\label{tab:reporting_checklist}
\end{table}

Overall, the current document is credible as a narrow technical report. It is
not yet a top-tier research paper because the strongest missing evidence is
comparative rather than editorial: controlled full-precision retention, official
judge scoring, runtime ablations, quantization ablations, and broader
replication.

\section{Threats to Validity}

\paragraph{Single model and quantization path.}
The main results use \qwenawq{} with AWQ Marlin. Other models,
quantization formats, and kernels may behave differently.

\paragraph{Fixed workload.}
The primary workload has approximately 512 input tokens and fixed 256 output
tokens. Real traffic varies in prompt length, output length, retry behavior,
tool calls, and burstiness.

\paragraph{Short-context throughput profile.}
The highest-throughput results use \texttt{max\_model\_len=1024}. They should
not be generalized to high-concurrency 8K, 16K, or 32K serving.

\paragraph{No BF16/FP16 retention yet.}
Absolute AWQ quality scores do not prove retention against full precision. A
matched baseline endpoint is required.

\paragraph{MT-Bench generation is not judge scoring.}
The MT-Bench run generated 80/80 answers, but no external judge score is
available.

\paragraph{Energy scope.}
Energy is GPU-board energy from \texttt{nvidia-smi}, not wall energy.

\paragraph{Small repeated-run sample.}
The repeated fixed-shape CI uses three runs per condition. This is enough to
catch gross instability, but broader statistical confidence would require more
runs across days, hosts, driver versions, and ambient conditions.

\section{Reproducibility}

The benchmark code and documentation are available at:

\begin{quote}
\url{https://github.com/Kevin-Li-2025/llm-quant-bench}
\end{quote}

Key remote artifact paths:

\begin{itemize}
  \item 24h soak:
        \texttt{/home/USER/llm-quant-bench/runs/qwen72b-awq-l20/}\\
        \texttt{soak-24h-fixed512x256-c10-20260519T034856Z}
  \item Repeated fixed-shape CI:
        \texttt{/home/USER/llm-quant-bench/runs/qwen72b-awq-l20/}\\
        \texttt{repeated-fixed512x256-vllm-awq-20260520T120138Z}
  \item Quality evaluation:
        \texttt{/home/USER/llm-quant-bench/runs/qwen72b-awq-l20/}\\
        \texttt{quality-eval-20260520T084323Z}
  \item BF16/MT-Bench preflight:
        \texttt{/home/USER/llm-quant-bench/runs/qwen72b-awq-l20/}\\
        \texttt{bf16-mtbench-preflight-20260520T123916Z}
\end{itemize}

Representative load command:

\begin{verbatim}
python3 -m llm_quant_bench load \
  --config runs/qwen72b-awq-l20/config.fixed512x256.json \
  --dataset runs/qwen72b-awq-l20/fixed-512in-256out.jsonl \
  --out runs/qwen72b-awq-l20/soak-24h-fixed512x256-c10-.../load \
  --concurrency 10 \
  --duration-seconds 86400 \
  --stream-usage
\end{verbatim}

Representative repeated-run command:

\begin{verbatim}
python3 scripts/run_repeated_load.py \
  --config runs/qwen72b-awq-l20/config.fixed512x256.json \
  --dataset runs/qwen72b-awq-l20/fixed-512in-256out.jsonl \
  --out runs/qwen72b-awq-l20/repeated-fixed512x256-vllm-awq \
  --concurrencies 1 4 8 16 \
  --repeats 3 \
  --duration-seconds 120 \
  --stream-usage
\end{verbatim}

\section{Conclusion}

This report shows that a single NVIDIA L20 can sustain \qwenawq{}
serving under a fixed short-context throughput workload. The strongest evidence
is the combination of a 24-hour concurrency-10 soak at 108.84 output tokens/s,
12/12 successful repeated fixed-shape runs, and a repeated concurrency-16 mean
of 127.22 $\pm$ 12.68 output tokens/s. The result is meaningful because it
documents a practical single-accelerator deployment point for a 70B-class model
with explicit latency, energy, quality, and limitation boundaries.

The correct interpretation is narrow but useful: single-L20 70B-class AWQ
serving is feasible and stable for a controlled throughput-oriented workload.
Stronger
claims about lossless quality, full-precision retention, judge-scored
instruction following, long-context high-concurrency serving, or general
production replacement require additional baseline and ablation experiments.

\bibliographystyle{plain}
\bibliography{references}

\end{document}